\documentclass[%
prl,
 reprint,
 amsmath,amssymb,amsfonts
 aps,preprintnumbers]{revtex4-1}

\usepackage[colorlinks=true,linkcolor=black, citecolor=black,
urlcolor=black]{hyperref}

\usepackage{multirow,graphics}
\usepackage{enumerate}
\usepackage{amstext}
\usepackage{amssymb}
\usepackage{amsmath}
\usepackage{graphicx}
\usepackage{color}
\usepackage{tikz}
\usepackage[nodisplayskipstretch]{setspace}
\usepackage{tcolorbox}
\usepackage{float}
\usepackage{wasysym}

\begin{document}

\newcommand{\Tr}{{\text{Tr}}}
\newcommand{\como}[1]{{\color[rgb]{0.0,0.1,0.9} {\bf \"O:} #1} }
\renewcommand{\d}{\mathrm{d}}

\makeatletter
\newcommand{\subalign}[1]{%
  \vcenter{%
    \Let@ \restore@math@cr \default@tag
    \baselineskip\fontdimen10 \scriptfont\tw@
    \advance\baselineskip\fontdimen12 \scriptfont\tw@
    \lineskip\thr@@\fontdimen8 \scriptfont\thr@@
    \lineskiplimit\lineskip
    \ialign{\hfil$\m@th\scriptstyle##$&$\m@th\scriptstyle{}##$\crcr
      #1\crcr
    }%
  }
}
\makeatother

\makeatletter
     \@ifundefined{usebibtex}{\newcommand{\ifbibtexelse}[2]{#2}} {\newcommand{\ifbibtexelse}[2]{#1}}
\makeatother


\usetikzlibrary{decorations.pathmorphing}
\usetikzlibrary{decorations.markings}
\usetikzlibrary{intersections}
\usetikzlibrary{calc}
  \usetikzlibrary{positioning}

\tikzset{
photon/.style={decorate, decoration={snake}},
particle/.style={postaction={decorate},
    decoration={markings,mark=at position .5 with {\arrow{>}}}},
antiparticle/.style={postaction={decorate},
    decoration={markings,mark=at position .5 with {\arrow{<}}}},
gluon/.style={decorate, decoration={coil,amplitude=2pt, segment length=4pt},color=purple},
wilson/.style={color=blue, thick},
scalarZ/.style={postaction={decorate},decoration={markings, mark=at position .5 with{\arrow[scale=1]{stealth}}}},
scalarX/.style={postaction={decorate}, dashed, dash pattern = on 4pt off 2pt, dash phase = 2pt, decoration={markings, mark=at position .53 with{\arrow[scale=1]{stealth}}}},
scalarZw/.style={postaction={decorate},decoration={markings, mark=at position .75 with{\arrow[scale=1]{stealth}}}},
scalarXw/.style={postaction={decorate}, dashed, dash pattern = on 4pt off 2pt, dash phase = 2pt, decoration={markings, mark=at position .60 with{\arrow[scale=1]{stealth}}}}
}


\newcommand{\footnoteab}[2]{\ifbibtexelse{%
\footnotetext{#1}%
\footnotetext{#2}%
\cite{Note1,Note2}%
}{%
\newcommand{\textfootnotea}{#1}%
\newcommand{\textfootnoteab}{#2}%
\cite{thefootnotea,thefootnoteab}}}

\def\e{\epsilon}
     \def\bT{{\bf T}}
    \def\bQ{{\bf Q}}
    \def\wT{{\mathbb{T}}}
    \def\wQ{{\mathbb{Q}}}
    \def\ttQ{{\bar Q}}
    \def\tQ{{\tilde \bP}}
        \def\bP{{\bf P}}
        \def\dq{{\dot q}}
    \def\CF{{\cal F}}
    \def\cC{\CF}
    
     \def\l{\lambda}
\def\hbZ{{\widehat{ Z}}}
\def\bZ{{\resizebox{0.28cm}{0.33cm}{$\hspace{0.03cm}\check {\hspace{-0.03cm}\resizebox{0.14cm}{0.18cm}{$Z$}}$}}}

\newcommand{\as}[2]{
  $a_{#1#2}$
}

\title{Cluster adjacency properties of scattering amplitudes}

\author{James Drummond$^{a}$, Jack Foster$^{a}$, \"Omer G\"urdo\u gan$^{a}$, }

\affiliation{%
  \(^{a}\)School of Physics \& Astronomy, University of Southampton,
  Highfield, Southampton, SO17 1BJ, United Kingdom.
}

\begin{abstract}
We conjecture a new set of analytic relations for scattering amplitudes in planar $\mathcal{N}=4$ super Yang-Mills theory. They generalise the Steinmann relations and are expressed in terms of the cluster algebras associated to $\mathrm{Gr}(4,n)$. In terms of the symbol, they dictate which letters can appear consecutively. We study heptagon amplitudes and integrals in detail and present symbols for previously unknown integrals at two and three loops which support our conjecture.
\end{abstract}

 \maketitle

\section{Introduction}

Scattering amplitudes in perturbative quantum field theory have an intricate analytic structure. In certain cases the analytic structure becomes tractable enough to submit to a general mathematical description. The case of scattering amplitudes in planar maximally supersymmetric Yang-Mills theory is particularly striking. Here there is evidence that the branch cut singularities of certain perturbative loop amplitudes are described by cluster algebras associated to the Grassmannians $\mathrm{Gr}(4,n)$ \cite{Golden:2013xva}. More precisely, the set of locations of the branch point singularities (the \emph{alphabet}) is described by the set of all cluster $\mathcal{A}$-coordinates of the associated cluster algebra. 

The $\mathcal{A}$-coordinates are polynomials in the Grassmannian Pl\"ucker coordinates $(ijkl)$ which are totally antisymmetric $sl_4$-invariant combinations of momentum twistors $Z_i$ \cite{Hodges:2009hk}. Such variables neatly describe the geometry of a polygonal light-like Wilson loop dual to the planar scattering amplitude \cite{Alday:2007hr,Drummond:2007aua,Brandhuber:2007yx}. The sides of the polygon are given by the massless particle momenta $p_i$ and the corners, denoted by $x_i$, are then related to lines in momentum twistor space $\mathbb{P}^3$ after picking a preferred bitwistor $I$, corresponding to the null cone at infinity,
\begin{equation}
p_i = x_{i+1} - x_i\,,\qquad x_i = \frac{Z_{i-1} {\scriptstyle \wedge} Z_i}{(i-1\,i\,I)}\,.
\end{equation}

Steinmann relations \cite{Steinmann,Steinmann2} are the requirement that a scattering amplitude does not have consecutive discontinuities in overlapping channels. In the context of amplitudes in planar $\mathcal{N}=4$ super yang-Mills theory their importance was emphasised in \cite{Bartels:2008sc} and they have been usefully employed to construct amplitudes in \cite{Caron-Huot:2016owq,Dixon:2016nkn}. 

In order to see the appearance of Steinmann relations in massless amplitudes it is useful to define an appropriate infrared finite quantity \cite{Caron-Huot:2016owq}. In planar $\mathcal{N}=4$ this quantity is the \emph{BDS-like subtracted amplitude}  \cite{Alday:2009dv} which exists for $n$-point amplitudes with $n\geq6$ and $n \neq 0$ mod 4. These amplitudes do not have consecutive branch cuts in overlapping three-particle or higher Mandelstam invariants. For example, a discontinuity around $s_{123}=0$ cannot itself have a discontinuity around $s_{234}=0$.

\section{Adjacency rules from ${\mathrm{ Gr}} (4,n)$ clusters}

{\bf Polylogarithms and symbols.}
The amplitudes that we discuss here are believed to be expressible in terms of polylogarithms, i.e. iterated integrals over a set of logarithmic singularities.
A polylogarithm (or `pure' function) of weight $k$, denoted as $f^{(k)}$,
satisfies
\begin{equation}
  \d\, f^{(k)}
  =
  \sum_r f^{(k-1)}_r\d \log \phi_r\, ,
\label{KZeq}
\end{equation}
where $\phi_r$ (the \emph{letters}) are algebraic functions of some set of coordinates. Functions of weight one are linear combinations of logarithms of the letters. The symbol map, defined following (\ref{KZeq})  as
\begin{equation}
  \mathcal S \bigl[f^{(k)}\bigr]
  =
  \sum_r \mathcal S [f^{(k-1)}_r] {\scriptstyle \otimes}\,  \phi_r\, ,
\end{equation}
maps a weight-$k$ function to a $k$-fold
tensor
\begin{equation}
  \label{eq:symbol}
  \sum_{\vec{\phi}} c_{\vec{\phi }}\,\,
  \phi_{r_1} {\scriptstyle \otimes} \, \phi_{r_2} {\scriptstyle \otimes} \dotsc {\scriptstyle \otimes} \, \phi_{r_k}\,,
\end{equation}
with rational coefficients $c_{\vec{\phi}}$, which encodes both the differential and the branch-cut structure of $f^{(k)}$.  In
order for an expression of the form (\ref{eq:symbol}) to be the symbol of
a function it has to satisfy the integrability condition,
\begin{equation}
  \sum_{\vec{\phi}} c_{\vec{\phi }}\,\,
  \phi_{r_1} {\scriptstyle \!\otimes}\,.\,.\,.\, \phi_{r_{i-1}} {\scriptstyle \! \otimes\,} \phi_{r_{i+2}} \ldots \phi_{r_k}
  \frac{\d \phi_{r_i} {\scriptstyle \wedge}\, \d\phi_{r_{i+1}}}{ \phi_{r_i}\phi_{r_{i+1}}}=0.
\end{equation}
The symbol map was used in \cite{Goncharov:2010jf} to deduce a beautifully simple formula for the two-loop hexagon remainder \cite{Bern:2008ap,Drummond:2008aq,DelDuca:2009au}. 
Terms of the form
$\phi_{r_1} {\scriptstyle \otimes} \, \phi_{r_2} {\scriptstyle \otimes}\dotsc $ in the symbol of a function $f^{(k)}$ indicate a logarithmic branch point
where  $\phi_{r_1}$ vanishes, while the discontinuity
across the associated branch cut is a function of weight $(k-1)$ whose symbol is the sum of such terms with the initial entry $\phi_{r_1}$ removed. Consequently, Steinmann relations which impose constraints on double discontinuities have implications on the initial pairs of entries of the symbol.

From (\ref{KZeq}), differential operators also have a simple action on the symbol, namely:
\begin{equation}
  \d (\phi_1 {\scriptstyle \otimes}\dotsc {\scriptstyle \otimes}\,\phi_k)
  =
\d \log \phi_k
  \bigl(\phi_1 {\scriptstyle \otimes} \dotsc {\scriptstyle \otimes}\,\phi_{k-1} \bigr)\,,
\end{equation}
allowing differential equations for integrals to be set up and solved
at the level of the symbol.

{\bf Cluster algebras.}
There is evidence \cite{Golden:2013xva} that cluster algebras \cite{1021.16017,1054.17024,Scott} address the question of which set of letters
$\phi_i$, called the \emph{alphabet}, appear in the symbol of certain scattering amplitudes in planar ${\cal N}=4$ super
Yang-Mills. A $\mathrm{Gr}(4,n)$ cluster algebra consists of clusters
that are represented by quiver diagrams that consist of \emph{nodes}
that are connected by arrows and carry the ${\cal A}$-coordinates,
which are the letters potentially appearing in the symbol of the
$n$-particle scattering amplitude.

For $\mathrm{Gr}(4,7)$, the so-called initial cluster is represented
by the following quiver diagram with Pl\"ucker coordinates at each of the nodes: 
{\small
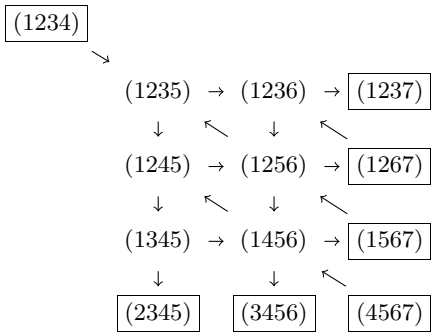
\begin{center}
\makeatletter  
\newcommand{\phantombox}[1]{%
  \setbox0=\hbox{#1}%
  \begin{tcolorbox}[colframe=white,colback=white,boxrule=0.4pt,
    left=2pt,right=2pt,top=3pt,bottom=3pt,boxsep=0pt,width=1.1cm, valign = center,  halign=center, sharp corners = all]
    #1
  \end{tcolorbox}
}
\newcommand{\frozenbox}[1]{%
  \setbox0=\hbox{#1}%
  \begin{tcolorbox}[colframe=black,colback=white,boxrule=0.5pt,
      left=2pt,right=2pt,top=2pt,bottom=2pt,boxsep=0pt,width=1.1cm, halign=center, sharp corners = all]
    #1
  \end{tcolorbox}
}
\makeatother
\begin{tikzpicture}%
    [
    unfrozen/.style={},
    frozen/.style={inner sep=1.2mm,outer sep=0mm,yshift=0},
    node distance = 0.2cm
    ]
    \node[frozen]        (f0) at (0,5) {$\frozenbox{(1234)}$};
    \node[frozen, below right = of f0]        (t1)  {$\phantombox{(1235)}$};
    \node[frozen, right = of t1]        (t2)  {$\phantombox{(1236)}$};
    \node[frozen, below = of t1]        (m1)  {$\phantombox{(1245)}$};
    \node[frozen, below = of t2]        (m2)  {$\phantombox{(1256)}$};
    \node[frozen, below = of m1]        (b1)  {$\phantombox{(1345)}$};
    \node[frozen, below = of m2]        (b2)  {$\phantombox{(1456)}$};
    \node[frozen, right = of t2]        (f1)  {$\frozenbox{(1237)}$};
    \node[frozen, right = of m2]        (f2)  {$\frozenbox{(1267)}$};
    \node[frozen, right = of b2]        (f3)  {$\frozenbox{(1567)}$};
    \node[frozen, below = of b1]        (f4)  {$\frozenbox{(2345)}$};
    \node[frozen, below = of b2]        (f5)  {$\frozenbox{(3456)}$};
    \node[frozen, right= of f5]         (f6)  {$\frozenbox{(4567)}$};
    
    \draw[->] (f0) -- (t1);
    \draw[->] (t1) -- (t2);    \draw[->] (t1) -- (m1) ;    \draw[->] (t2) -- (f1) ;  \draw[->] (t2) -- (m2) ;
    \draw[->] (m1) -- (m2);  \draw[->] (m1) -- (b1) ; \draw[->] (m2) -- (t1);      \draw[->] (m2) -- (f2) ;  \draw[->] (m2) -- (b2) ; \draw[->] (f2) -- (t2);
    \draw[->] (b1) -- (b2);  \draw[->] (b1) -- (f4) ; \draw[->] (b2) -- (m1);      \draw[->] (b2) -- (f3) ;  \draw[->] (b2) -- (f5) ; \draw[->] (f3) -- (m2);
    \draw[->] (f6) -- (b2);
\end{tikzpicture}
\end{center}}
The $\mathrm{Gr}(4,6)$ initial cluster is obtained from the above by deleting the final column and placing boxes around the unboxed nodes in the second column. The boxed nodes are referred to as \emph{frozen} nodes and the rest as \emph{unfrozen}.

A cluster is mapped to a different one under an operation called
\emph{mutation} which replaces the value of an unfrozen node and changes the
connectivity of the quiver according to the rules explained below.

Let a quiver with nodes $\{a_i\}$ be described by the antisymmetric adjacency matrix
$b_{ij}$ defined as
\begin{equation}
  b_{ij} = (\# \text{arrows } i\to j) - (\# \text{arrows } j\to i)\, .
\end{equation}
If the node $a_k$ of a cluster is mutated then the adjacency matrix of
the resulting cluster is
\begin{equation}
  b'_{ij} =
  \begin{cases}
    -b_{ij}                  &   k \in \{i,j\}\\
    b_{ij}                   &   b_{ik}b_{kj} \leq 0\\
    b_{ij}  + b_{ik}b_{kj}    &   b_{ik}b_{kj} > 0\\
    b_{ij}  - b_{ik}b_{kj}    &   b_{ik}b_{kj} < 0\\
  \end{cases}
\end{equation}
and the value of the node mutates to 
\begin{equation}
  a_k \mapsto \frac{1}{a_k}\,\,
  \biggl[
  \prod_{i|b_{ik}>0} a_{i}^{b_{ik}} +   \prod_{i|b_{ik}<0} a_{i}^{-b_{ik}} 
  \biggr]\,.
\end{equation}

Mutations define a connectivity structure between different
clusters, such that each cluster can be thought of living on a vertex
of a polytope. For $n\geq8$ the mutations do not close on
finitely many clusters and therefore the polytope has infinitely many vertices.

The ${\rm Gr}(4,7)$ cluster algebra generates 49 distinct ${\cal A}$
coordinates, distributed in sextets across 833 clusters. These 49
coordinates admit 42 homogeneous ratios which can be chosen as \cite{Drummond:2014ffa}
\begin{equation}
\begin{aligned}[b]
  a_{11} &= \frac{(1234)(1567)(2367)}{(1237)(1267)(3456)}\\
  a_{31} &= \frac{(1567)(2347)}{(1237)(4567)}\\
  a_{51} &= \frac{(1(23)(45)(67))}{(1234)(1567)}
\end{aligned}
\,\,\,
\begin{aligned}[b]
  a_{21} &= \frac{(1234)(2567)}{(1267)(2345)}\\
  a_{41} &= \frac{(2357)(3456)}{(2345)(4567)}  \\
  a_{61} &= \frac{(1(34)(56)(72))}{(1234)(1567)}
\end{aligned}\,
\label{heptletters}
\end{equation}
and cyclically related $a_{ij}$ for $j>1$. Here we write {\small $(a(bc)(de)(fg)) =(acde)(acfg)-(acde)(abfg)$}. For $\mathrm{Gr}(4,6)$ there 14 clusters generating 15 $\mathcal{A}$-coordinates admitting 9 homogeneous ratios.
Note that for any $n$ odd, there is a one-to-one map between the
$\cal A$-coordinates and the homogeneous letters of the alphabet in
which the frozen nodes, which are Pl\"ucker coordinates with adjacent
indices, are effectively set to 1. Therefore we can assign one of the
$a_{ij}$ for each unfrozen node of a $\mathrm{Gr}(4,7)$
cluster so that the initial cluster effectively becomes:  
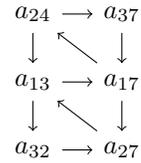
\begin{center}
\begin{tikzpicture}%
    [
    unfrozen/.style={},
    frozen/.style={inner sep=1.5mm,outer sep=0mm,yshift=0},
    node distance = 0.4cm
    ]
    \node[frozen]  (t1)  at (1,4) {$a_{24}$};
    \node[frozen, right = of t1]        (t2)  {$a_{37}$};
    \node[frozen, below = of t1]        (m1)  {$a_{13}$};
    \node[frozen, below = of t2]        (m2)  {$a_{17}$};
    \node[frozen, below = of m1]        (b1)  {$a_{32}$};
    \node[frozen, below = of m2]        (b2)  {$a_{27}$};
    
    \draw[->] (t1) -- (t2); \draw[->] (t1) -- (m1); \draw[->] (t2) -- (m2);
    \draw[->] (m1) -- (m2); \draw[->] (m1) -- (b1); \draw[->] (m2) -- (t1); \draw[->] (m2) -- (b2) ;
    \draw[->] (b1) -- (b2); \draw[->] (b2) -- (m1);      

  \end{tikzpicture}\,.
\end{center}

It is possible to mutate the initial cluster on nodes $a_{17}$, $a_{27}$ and $a_{32}$ to reach a cluster that has the topology of an $E_6$ Dynkin diagram:
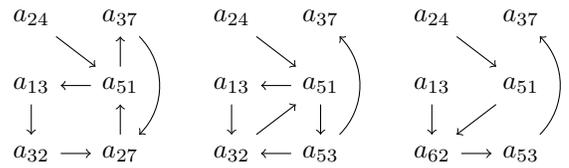
\begin{center}
  \begin{tikzpicture}%
    [
    unfrozen/.style={},
    frozen/.style={inner sep=1.5mm,outer sep=0mm,yshift=0},
    node distance = 0.4cm
    ]
    \node[frozen]  (t1)  at (1,4) {$a_{24}$};
    \node[frozen, right = of t1]        (t2)  {$a_{37}$};
    \node[frozen, below = of t1]        (m1)  {$a_{13}$};
    \node[frozen, below = of t2]        (m2)  {$a_{51}$};
    \node[frozen, below = of m1]        (b1)  {$a_{32}$};
    \node[frozen, below = of m2]        (b2)  {$a_{27}$};
    
    \draw[->] (m2) -- (t2);
    \draw[->] (m2) -- (m1); \draw[->] (m1) -- (b1); \draw[->] (t1) -- (m2); \draw[->] (b2) -- (m2) ;
    \draw[->] (b1) -- (b2);
    \draw[->] (t2) to[out=-45, in = 45] (b2);

\end{tikzpicture}
\quad
  \begin{tikzpicture}%
    [
    unfrozen/.style={},
    frozen/.style={inner sep=1.5mm,outer sep=0mm,yshift=0},
    node distance = 0.4cm
    ]
    \node[frozen]  (t1)  at (1,4) {$a_{24}$};
    \node[frozen, right = of t1]        (t2)  {$a_{37}$};
    \node[frozen, below = of t1]        (t)  {$a_{13}$};
    \node[frozen, below = of t2]        (m2)  {$a_{51}$};
    \node[frozen, below = of m1]        (b1)  {$a_{32}$};
    \node[frozen, below = of m2]        (b2)  {$a_{53}$};
    
    \draw[->] (m2) -- (m1); \draw[->] (m1) -- (b1); \draw[->] (t1) -- (m2); \draw[->] (m2) -- (b2) ;
    \draw[->] (b2) -- (b1);
    \draw[->] (b1) -- (m2);
    \draw[->] (b2) to[out=45, in = -45] (t2);

\end{tikzpicture}
\quad
  \begin{tikzpicture}%
    [
    unfrozen/.style={},
    frozen/.style={inner sep=1.5mm,outer sep=0mm,yshift=0},
    node distance = 0.4cm
    ]
    \node[frozen]  (t1)  at (1,4) {$a_{24}$};
    \node[frozen, right = of t1]        (t2)  {$a_{37}$};
    \node[frozen, below = of t1]        (m1)  {$a_{13}$};
    \node[frozen, below = of t2]        (m2)  {$a_{51}$};
    \node[frozen, below = of m1]        (b1)  {$a_{62}$};
    \node[frozen, below = of m2]        (b2)  {$a_{53}$};
    
    \draw[->] (m1) -- (b1); \draw[->] (t1) -- (m2); 
    \draw[->] (b1) -- (b2);
    \draw[->] (m2) -- (b1);
    \draw[->] (b2) to[out=45, in = -45] (t2);\

\end{tikzpicture}\,.

\end{center}
Finally mutating $a_{37}$ to $a_{26}$, $a_{53}$ to $a_{41}$ and
$a_{26}$ to $a_{33}$ only changes the direction of some arrows and yields:
\begin{center}
  \begin{tikzpicture}%
    [
    unfrozen/.style={},
    frozen/.style={inner sep=1.5mm,outer sep=0mm,yshift=0},
    node distance = 0.4cm
    ]

    \node[frozen]        (t)  at (1,4) {$a_{13}$};
    \node[frozen, below = of t]        (b3)  {$a_{62}$};
       \node[frozen, left = of b3]        (b2)  {$a_{51}$};
         \node[frozen, left = of b2]  (b1)   {$a_{24}$};
    \node[frozen, right = of b3]        (b4)  {$a_{41}$};
       \node[frozen, right = of b4]        (b5)  {$a_{33}$};
    \draw[->] (t) -- (b3); \draw[->] (b1) -- (b2); 
    \draw[->] (b2) -- (b3);
    \draw[->] (b4) -- (b3);
    \draw[->] (b5) -- (b4);\
\end{tikzpicture}\,.
\end{center}
\vspace{-1.5cm}
\begin{equation}
\label{E6cluster}
\end{equation}
\vspace{0cm}

{\bf Cluster adjacency.}
While mutations of clusters generate the letters of the symbol
alphabet, the alphabet itself does not contain information on the details of which clusters contain which $\mathcal{A}$-coordinates nor which clusters are linked by mutations.
However, a survey of all
known MHV and NMHV BDS-like subtracted heptagon amplitudes reveals that only certain
pairs of letters appear in neighbouring slots. This leads us to
conjecture a much more general set of adjacency relations for BDS-like subtracted amplitudes: 
\vspace{0.1cm}
\begin{center}
\boxed{
\begin{minipage}{0.95\linewidth}
\emph{Two distinct $\cal{A}$-coordinates can appear consecutively in a symbol
  only if there exists a cluster where they both appear.}
\end{minipage}
}
\vspace{0.1cm}
\end{center}
We believe that the above conjecture will apply to any BDS-like subtracted amplitude which is expressed in terms of cluster polylogarithms. It has been conjectured that all MHV and NMHV amplitudes in planar $\mathcal{N}=4$ super Yang-Mills will have a polylogarithmic form \cite{ArkaniHamed:2012nw}, though it has not yet been directly tested whether the alphabets are dictated by the $\mathrm{Gr}(4,n)$ cluster structures for $n\geq8$ beyond two loops or beyond MHV amplitudes. 

For the eight-point amplitudes, a BDS-like subtracted amplitude in the sense of \cite{Alday:2009dv}, which uses only two-particle Mandelstam invariants to provide a solution to the conformal Ward identity of \cite{Drummond:2007au}, does not exist. 
Proceeding to nine points, one again has a canonical BDS-like subtracted amplitude, constructible from the two-loop results in \cite{CaronHuot:2011ky} and we have verified directly that it does obey the above conjecture. That is, for each neighbouring pair appearing in the symbol we can find a cluster containing that pair.

It is important to emphasise that the above adjacency conjecture introduces a much more detailed role for the cluster structure over and above the fact that the alphabet can be obtained from the union over all cluster $\mathcal{A}$-coordinates. It is the structure of the individual clusters which constrains both sequences of discontinuities (reading the symbol from the left) and successive derivatives (reading from the right).

In the case of planar $\mathcal{N}=4$ heptagon amplitudes there are 840 distinct admissible ordered pairs of letters out of the 1764 possible ordered pairs one can make from the 42 letters.
We summarise this information in Table
\ref{fig:adjacency}, where we also distinguish whether the pairs that
appear together in a cluster are connected by a quiver arrow as well as whether
two letters that never appear together mutate into each
other. If one letter mutates into another, this pair
never appear together in a third cluster. 

\newcommand{\bdiamond}[1][fill=black]{\tikz [x=1.2ex,y=1.2ex,line width=.1ex,line join=round, yshift=-0.285ex] \draw  [#1]  (0,0) -- (.5,.5) -- (0,1) -- (-0.5,0.5)  -- cycle;}%
\newcommand{\wdiamond}[1][fill=white]{\tikz [x=1.2ex,y=1.2ex,line width=.1ex,line join=round, yshift=-0.285ex] \draw  [#1]  (0,0) -- (.5,.5) -- (0,1) -- (-0.5,0.5)  -- cycle;}%
\newcommand{\bcircle}[1][fill=black]{\tikz [x=1.2ex,y=1.2ex,line width=.1ex,line join=round, yshift=-0.285ex] \draw  [#1]  (0,0) circle (0.5);}%
\newcommand{\wcircle}[1][fill=white]{\tikz [x=1.2ex,y=1.2ex,line width=.1ex,line join=round, yshift=-0.285ex] \draw  [#1]  (0,0) circle (0.5);}%

\begin{table*}
  \centering
  \begin{tikzpicture}
        \node (table) {  
\scalebox{0.85}{\begin{tabular}{r|ccccccc|ccccccc|ccccccc|ccccccc|ccccccc|ccccccc}
& \multicolumn{7}{|c}{\as{1}{i}} & \multicolumn{7}{|c}{\as{2}{i}} & \multicolumn{7}{|c}{\as{3}{i}} & \multicolumn{7}{|c}{\as{4}{i}} & \multicolumn{7}{|c}{\as{5}{i}} & \multicolumn{7}{|c}{\as{6}{i}} \\
\hline
\as{1}{1}& $\bcircle$& $\wcircle$& $\wcircle$& $\bdiamond$& $\bdiamond$& $\wcircle$& $\wcircle$& $\bdiamond$& $\bdiamond$& $\wcircle$& $\bcircle$& $\bdiamond$& $\bcircle$& $\wcircle$& $\bdiamond$& $\wcircle$& $\bcircle$& $\bdiamond$& $\bcircle$& $\wcircle$& $\bdiamond$& $\bcircle$& $\wcircle$& $\bdiamond$& $\wcircle$& $\wcircle$& $\bdiamond$& $\wcircle$& $\bcircle$& $\wcircle$& $\bdiamond$& $\wcircle$& $\wcircle$& $\bdiamond$& $\wcircle$& $\wdiamond$& $\bdiamond$& $\wcircle$& $\wcircle$&  $\wcircle$& $\wcircle$& $\bdiamond$\\
\as{2}{1}& $\bdiamond$& $\wcircle$& $\bcircle$& $\bdiamond$& $\bcircle$& $\wcircle$& $\bdiamond$& $\bcircle$& $\wcircle$& $\bcircle$& $\bdiamond$& $\bdiamond$& $\bcircle$&   $\wcircle$& $\bdiamond$& $\wcircle$& $\bdiamond$& $\bdiamond$& $\wcircle$& $\bdiamond$& $\bcircle$& $\bdiamond$& $\wcircle$& $\bdiamond$& $\wcircle$& $\bcircle$& $\bdiamond$& $\wcircle$& $\wcircle$& $\bdiamond$& $\bcircle$& $\wcircle$& $\bdiamond$& $\wcircle$& $\bdiamond$& $\wcircle$& $\bdiamond$& $\wcircle$& $\bcircle$& 
  $\wcircle$& $\bdiamond$& $\wcircle$\\
\as{3}{1}& $\bdiamond$& $\bdiamond$& $\wcircle$& $\bcircle$& $\bdiamond$& $\bcircle$& $\wcircle$& $\bdiamond$& $\bcircle$& $\bdiamond$& $\wcircle$& $\bdiamond$& $\bdiamond$&  $\wcircle$& $\bcircle$& $\wcircle$& $\bcircle$& $\bdiamond$& $\bdiamond$& $\bcircle$& $\wcircle$& $\bdiamond$& $\wcircle$& $\bdiamond$& $\bcircle$& $\wcircle$& $\bdiamond$& $\wcircle$& $\wcircle$& $\bdiamond$& $\wcircle$& $\bdiamond$& $\wcircle$& $\bcircle$& $\bdiamond$& $\wcircle$& $\wcircle$& $\bdiamond$& $\wcircle$& 
  $\bcircle$& $\wcircle$& $\bdiamond$\\
\as{4}{1}&$\bcircle$& $\wcircle$& $\bdiamond$& $\wcircle$& $\wcircle$& $\bdiamond$& $\wcircle$& $\bdiamond$& $\wcircle$& $\bdiamond$& $\bcircle$& $\wcircle$& $\bdiamond$& $\wcircle$& $\bdiamond$& $\wcircle$& $\bdiamond$& $\wcircle$& $\bcircle$& $\bdiamond$& $\wcircle$& $\bcircle$& $\wdiamond$& $\bdiamond$& $\wcircle$& $\wcircle$& $\bdiamond$& $\wdiamond$& $\bcircle$& $\wcircle$& $\wcircle$& $\wcircle$& $\wcircle$& $\wcircle$& $\wcircle$& $\wdiamond$& $\bdiamond$& $\wdiamond$& $\wcircle$& 
  $\wcircle$& $\wdiamond$& $\bdiamond$\\
\as{5}{1}& $\bcircle$& $\wcircle$& $\bdiamond$& $\wcircle$& $\wcircle$& $\bdiamond$& $\wcircle$& $\wcircle$& $\bdiamond$& $\wcircle$& $\bdiamond$& $\wcircle$& $\bcircle$& $\bdiamond$& $\wcircle$& $\bdiamond$& $\bcircle$& $\wcircle$& $\bdiamond$& $\wcircle$& $\bdiamond$& $\bcircle$& $\wcircle$& $\wcircle$& $\wcircle$& $\wcircle$& $\wcircle$& $\wcircle$& $\bcircle$& $\wdiamond$& $\bdiamond$& $\wcircle$& $\wcircle$& $\bdiamond$& $\wdiamond$& $\wdiamond$& $\bdiamond$& $\wdiamond$& $\wcircle$& $\wcircle$& $\wdiamond$& $\bdiamond$\\
\as{6}{1}& $\wdiamond$& $\bdiamond$& $\wcircle$& $\wcircle$& $\wcircle$& $\wcircle$& $\bdiamond$& $\wcircle$& $\wcircle$& $\bdiamond$& $\wcircle$& $\bcircle$& $\wcircle$& $\bdiamond$& $\wcircle$& $\bdiamond$& $\wcircle$& $\bcircle$& $\wcircle$& $\bdiamond$& $\wcircle$& $\wdiamond$& $\bdiamond$& $\wdiamond$& $\wcircle$& $\wcircle$& $\wdiamond$& $\bdiamond$& $\wdiamond$& $\bdiamond$& $\wdiamond$& $\wcircle$& $\wcircle$& $\wdiamond$& $\bdiamond$& $\bcircle$& $\wdiamond$& $\wcircle$& $\wdiamond$& $\wdiamond$& $\wcircle$& $\wdiamond$\\
\end{tabular}
}
};
        \draw [blue, dashed]
        (-5.77,0.69) rectangle (-3.83,0.96);
  \end{tikzpicture}
\caption{The neighbourhood and connectivity relations of the coordinates \as{i}{1} with the 42-letter alphabet. Other relations can be inferred by cyclic symmetry. The relations in the dashed box imply the Steinmann conditions. \\
$\bdiamond$: There are clusters where the coordinates appear together connected by an arrow.\,\,\\
$\bcircle$: There are clusters where the coordinates appear together but they are never connected.\,\, \\
$\wcircle$: The coordinates never appear in the same cluster but there is a mutation that links them.\,\,\\
$\wdiamond$: The coordinates do not appear in the same cluster nor there is a mutation that links them. \\
}
  \label{fig:adjacency}  
\end{table*}

The constraints on the symbol due to the Steinmann conditions mentioned above come as a special case of the
cluster adjacency conditions: The letters $a_{12}$, $a_{13}$, $a_{16}$
or $a_{17}$ never appear in the same cluster as the letter
$a_{11}$. In fact, in four of the dihedral copies of the initial
cluster, a mutation on $a_{11}$ generates a different cluster in which
it is replaced by one of these four letters that are not allowed to
appear after an initial $a_{11}$. The Steinmann condition as detailed in \cite{Caron-Huot:2016owq} only applies to the first two letters of the symbol. However it has been observed \cite{DP,Yorgosslides} that for the hexagon amplitudes the same condition applies everywhere in the symbol. The adjacency property we see is in accord with this observation.

Only 784 of the 840 allowed adjacencies actually occur in the known
7-particle amplitudes, while the pairs
  \begin{subequations}
    \begin{align}
      [a_{11} {\scriptstyle \otimes}\, a_{41}] &\qquad  \text{\& cyclic + parity}\label{clusterbutnotultra}\\
    [a_{21} {\scriptstyle \otimes}\, a_{64}] &\qquad  \text{\& cyclic + reflection}        
    \end{align}
  \end{subequations}
  and the reversed pairs do not appear even though they are permitted by
  our conjecture. Nevertheless, in the following section we compute
  the symbol of a three-loop integral by constraining the space of
  weight-six Steinmann functions and find that it is consistent
  with our conjecture and has adjacent pairs of the form
  (\ref{clusterbutnotultra}) in its symbol.

\section{Heptagon integrals}
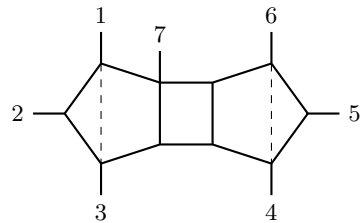
\begin{figure}[H]
\centering
\begin{tikzpicture}[scale=0.7]
    \newdimen\R
  \R=1cm
  \newdimen\L
  \L=0.6cm

  \coordinate (l1) at ($(36:\R) + (\R,0) + (-36:\R) + (72:\R)$);
  \coordinate (l2) at ($(36:\R) + (\R,0) + (-36:\R) + (0:\R)$);
  \coordinate (l3) at ($(36:\R) + (\R,0) + (-36:\R) + (-72:\R)$);

  \coordinate (c1) at (36:\R);
  \coordinate (c2) at (-36:\R) ;
  \coordinate (c3) at ($(36:\R) + (\R,0)$);
  \coordinate (c4) at ($(-36:\R) + (\R,0)$) ;

  \coordinate (r1) at (108:\R) ;
  \coordinate (r2) at (180:\R);
  \coordinate (r3) at (-108:\R);

    \draw[thick] (c1) -- (c2);
    \draw[thick] (c3) -- (c4);

    \draw[thick] (l1) -- (l2) -- (l3) -- (c4) -- (c2) -- (r3) -- (r2) -- (r1) -- (c1) -- (c3) -- (l1);
    \draw[dashed] (l1) -- (l3);
    \draw[dashed] (r1) -- (r3);

    \draw[thick] (l2) -- +(0:\L) node[at end, right] {$5$};
    \draw[thick] (r2) -- +(+180:\L) node[at end, left] {$2$};
    \draw[thick] (r1) -- +(0,\L) node[at end, above] {$1$};
    \draw[thick] (l1) -- +(0,\L) node[at end, above] {$6$};
    \draw[thick] (l3) -- +(0,-\L) node[at end, below] {$4$};
    \draw[thick] (r3) -- +(0,-\L) node[at end, below] {$3$};
    \draw[thick] (c1) -- +(90:\L) node[at end, above] {$7$};
    
\end{tikzpicture}
\caption{Seven-point, three-loop, massless integral.}
\label{fig:3loopint}
\end{figure}

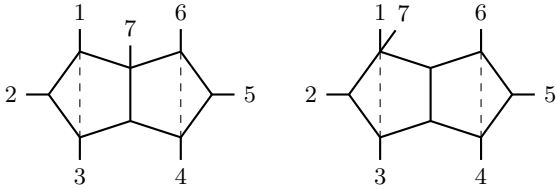
\begin{figure}[H]
\centering
\begin{tikzpicture}[scale=1]
  \newdimen\R
  \R=0.6cm
  \newdimen\L
  \L=0.3cm

  \coordinate (l1) at ($(36:\R) + (-36:\R) + (72:\R)$);
  \coordinate (l2) at ($(36:\R) + (-36:\R) + (0:\R)$);
  \coordinate (l3) at ($(36:\R) + (-36:\R) + (-72:\R)$);

  \coordinate (c1) at (36:\R);
  \coordinate (c2) at (-36:\R) ;

  \coordinate (r1) at (108:\R) ;
  \coordinate (r2) at (180:\R);
  \coordinate (r3) at (-108:\R);

  \draw[thick] (c1) -- (c2);
  \draw[thick] (l1) -- (l2) -- (l3) -- (c2) -- (r3) -- (r2) -- (r1) -- (c1) -- (l1);
    \draw[dashed] (l1) -- (l3);
    \draw[dashed] (r1) -- (r3);

    \draw[thick] (l2) -- +(0:\L) node[at end, right] {$5$};
    \draw[thick] (r2) -- +(+180:\L) node[at end, left] {$2$};
    \draw[thick] (r1) -- +(0,\L) node[at end, above] {$1$};
    \draw[thick] (l1) -- +(0,\L) node[at end, above] {$6$};
    \draw[thick] (l3) -- +(0,-\L) node[at end, below] {$4$};
    \draw[thick] (r3) -- +(0,-\L) node[at end, below] {$3$};
    \draw[thick] (c1) -- +(90:\L) node[at end, above] {$7$};
  \end{tikzpicture}
  \quad
  \begin{tikzpicture}[scale=1]
      \newdimen\R
      \R=0.6cm
      \newdimen\L
      \L=0.3cm

  \coordinate (l1) at ($(36:\R) + (-36:\R) + (72:\R)$);
  \coordinate (l2) at ($(36:\R) + (-36:\R) + (0:\R)$);
  \coordinate (l3) at ($(36:\R) + (-36:\R) + (-72:\R)$);

  \coordinate (c1) at (36:\R);
  \coordinate (c2) at (-36:\R) ;

  \coordinate (r1) at (108:\R) ;
  \coordinate (r2) at (180:\R);
  \coordinate (r3) at (-108:\R);

  \draw[thick] (c1) -- (c2);
  \draw[thick] (l1) -- (l2) -- (l3) -- (c2) -- (r3) -- (r2) -- (r1) -- (c1) -- (l1);
    \draw[dashed] (l1) -- (l3);
    \draw[dashed] (r1) -- (r3);

    \draw[thick] (l2) -- +(0:\L) node[at end, right] {$5$};
    \draw[thick] (r2) -- +(+180:\L) node[at end, left] {$2$};
    \draw[thick] (r1) -- +(90:\L) node[at end, above] {$1$};
    \draw[thick] (l1) -- +(0,\L) node[at end, above] {$6$};
    \draw[thick] (l3) -- +(0,-\L) node[at end, below] {$4$};
    \draw[thick] (r3) -- +(0,-\L) node[at end, below] {$3$};
    \draw[thick] (r1) -- +(54:0.35cm) node[at end, above] {$\,\,\,\,7$};
  \end{tikzpicture}
  \caption{The seven-point, two-loop integrals $I^{(2)}$ and $\tilde{I}^{(2)}$.}
  \label{fig:2loopint}
\end{figure}

Let us consider the following seven-point three-loop double pentaladder (drawn in Fig. \ref{fig:3loopint})
\begin{align}
\begin{split}
I^{(3)} = &\int d\tilde{Z} \frac{1}{\prod_{i=1}^{4}(AB\,i-1\,i)\prod_{i=4}^{7}(EF\,i-1\,i)} \\ &\times \frac{(AB13)(EF46)N}{(CD34)(ABCD)(CDEF)(CD67)}\,.
\end{split}
\end{align}
The measure is $d\tilde{Z}=\frac{d^4 Z_{AB}}{i \pi^2} \frac{d^4 Z_{CD}}{i \pi^2} \frac{d^4 Z_{EF}}{i \pi^2}$ and the numerator is $N = (2345)(3467)(7 (12)(34)(56))$ which ensures that $I^{(3)}$ is finite and has unit leading singularity \cite{ArkaniHamed:2010gh}. 
There exists a set of four second-order differential operators which relate $I^{(3)}$ to two-loop integrals \cite{Drummond:2010cz}. Using the notation $O_{ij} = Z_i \cdot \frac{\partial}{\partial Z_j}$ they are 
\begin{align}
(4567) N O_{45}O_{34} N^{-1} I^{(3)} &= -(3467) I^{(2)}\,, \label{diffeq1}\\
(3456) N O_{65}O_{76} N^{-1} I^{(3)} &= -(3467) I^{(2)}\,, \label{diffeq2}\\
(1237) N O_{32}O_{43} N^{-1} I^{(3)} &= +(1347) \tilde{I}^{(2)}\,, \label{diffeq3}\\
(1234) N O_{12}O_{71} N^{-1} I^{(3)} &= -(1347) \tilde{I}^{(2)}\,. \label{diffeq4}
\end{align}
The two-loop integrals $I^{(2)}$ and $\tilde{I}^{(2)}$ are shown in Fig. \ref{fig:2loopint}.

The second-order operators above reduce the weight by two, therefore they must annihilate the final entries in the symbol of $I^{(3)}$. Using this condition we can construct a set of ten multiplicative combinations out the 42 possible heptagon letters for the final entries.

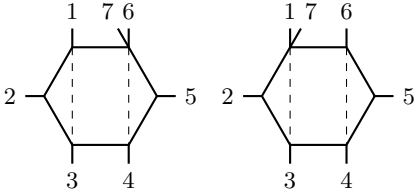
\begin{figure}[H]
\centering
\begin{tikzpicture}[scale=0.5]
	\newdimen\R
	\R=1.5cm
	\draw[thick] (0:\R) \foreach \x in {60,120,...,360} {  -- (\x:\R) };
	\draw[dashed] (60:\R) -- (300:\R);
	\draw[dashed] (120:\R) -- (240:\R);
	\draw[thick] (60:\R) -- +(90:0.5) node[at end, above] {$6$};
	\draw[thick] (120:\R) -- +(0,0.5) node[at end, above] {$1$};
	\draw[thick] (60:\R) -- +(+120:0.575) node[at end, above] {$\!\!\!\!\!7$};
	\draw[thick] (180:\R) -- +(+180:0.5) node[at end, left] {$2$};
	\draw[thick] (240:\R) -- +(0,-0.5) node[at end, below] {$3$};
	\draw[thick] (300:\R) -- +(0,-0.5) node[at end, below] {$4$};
	\draw[thick] (360:\R) -- +(+180:-0.5) node[at end, right] {$5$};
\end{tikzpicture}
\begin{tikzpicture}[scale=0.5]
	\newdimen\R
	\R=1.5cm
	\draw[thick] (0:\R) \foreach \x in {60,120,...,360} {  -- (\x:\R) };
	\draw[dashed] (60:\R) -- (300:\R);
	\draw[dashed] (120:\R) -- (240:\R);	
	\draw[thick] (60:\R) -- +(0,0.5) node[at end, above] {$6$};
	\draw[thick] (120:\R) -- +(90:0.5) node[at end, above] {$1$};
	\draw[thick] (120:\R) -- +(+60:0.575) node[at end, above] {$\,\,\,\,\,7$};
	\draw[thick] (180:\R) -- +(+180:0.5) node[at end, left] {$2$};
	\draw[thick] (240:\R) -- +(0,-0.5) node[at end, below] {$3$};
	\draw[thick] (300:\R) -- +(0,-0.5) node[at end, below] {$4$};
	\draw[thick] (360:\R) -- +(+180:-0.5) node[at end, right] {$5$};
\end{tikzpicture}
 \caption{The one-loop hexagon integrals $I^{(1)}$ and $\tilde{I}^{(1)}$.}
\label{fig:1loopint}
\end{figure}

The integral $I^{(2)}$ obeys a similar set of differential equations, except that the integrals on the RHS are the one-loop hexagons $I^{(1)}$ and $\tilde{I}^{(1)}$ depicted in Fig. \ref{fig:1loopint}.
Therefore the integral $I^{(2)}$ can also only have the same ten possible final entries as $I^{(3)}$.
Of the 322 weight-4 Steinmann heptagon symbols constructed in \cite{Dixon:2016nkn}, there is a unique linear combination with the correct final entries for $I^{(2)}$ which we conclude must be the result up to a scale. The details of the RHS differential equations were not required to obtain it, and indeed they can be used to derive formulas for the one-loop hexagons in Fig. \ref{fig:1loopint}.

Returning to $I^{(3)}$, we find that of the 3192 weight-6 Steinmann heptagon symbols constructed in \cite{Dixon:2016nkn}, seven of them have good final entries. However, only one of these produces our result for $I^{(2)}$ on the RHS of equations (\ref{diffeq1}) and (\ref{diffeq2}) and we conclude this is the symbol of $I^{(3)}$. Either of the equations (\ref{diffeq3}) or (\ref{diffeq4}) can then be used to derive a result for $\tilde{I}^{(2)}$. We quote all symbols in a file attached to the arXiv submission.

\indent When analysing the symbol of the three-loop integral $I^{(3)}$, pairs not present in the MHV and NMHV heptagon data were found, namely $[a_{11} {\scriptstyle \otimes}\, a_{41}]$ and $[a_{41} {\scriptstyle \otimes}\, a_{11}]$. Their cyclic and parity copies will therefore be found in the associated cyclic and parity copies of the integral, completing the set (\ref{clusterbutnotultra}). This evidence supports our conjecture that it is the cluster structure which controls the appearance of consecutive letters.

\section{Consequences of the adjacency rules}

{\bf Neighbour sets.}
Let us describe the sets of allowed neighbours of a given letter. For this the $E_6$ topology cluster depicted in (\ref{E6cluster}) is very useful. It contains one of each of the six cyclic classes of letters given in (\ref{heptletters}).  If we freeze the node $a_{13}$ the cluster algebra reduces to an $A_5$ algebra which generates 20 letters all of which are in a cluster with $a_{13}$. Including $a_{13}$ itself we find 21 possible neighbours for $a_{13}$. The letters $a_{16}$ and $a_{17}$ are in the neighbour set of $a_{13}$ but $a_{11}$, $a_{12}$, $a_{14}$ and $a_{15}$ are not, which implies the Steinmann conditions. The analysis applies similarly to all $a_{1i}$ type letters and is in accordance with the first line of Table \ref{fig:adjacency}. If we freeze either of the nodes $a_{24}$ or $a_{37}$ the cluster algebra reduces to a $D_5$ algebra, generating 25 allowed neighbours in addition to the letter itself. Likewise freezing either $a_{41}$ or $a_{51}$ leads to an $A_4 \times A_1$ algebra which generates $14+2=16$ neighbouring letters in addition to the letter itself. Finally if we freeze the node $a_{62}$ we obtain an $A_2 \times A_2 \times A_1$ subalgebra which generates $5+5+2=12$ allowed neighbours in addition to $a_{62}$ itself. Each subalgebra responsible for generating the allowed neighbour set of a given letter corresponds to a subpolytope in the whole $E_6$ polytope.

{\bf Integrability.}
Let us now analyse the restrictions that integrability places on symbols. We may construct all integrable weight-two symbols which obey the cluster adjacency conditions. In the heptagon case we find 573 possible weight two integrable words (with no additional conditions on the initial entries). Of these 441 are symmetric in the exchange of the initial and final entries (and hence are trivially integrable) and 132 are antisymmetric. The symmetric ones correspond to products of logarithms and the antisymmetric ones to combinations of dilogarithms. 

The number of antisymmetric symbols is the same as without the cluster adjacency conditions. Indeed, inspecting the pairs which appear in the antisymmetric integrable weight-two words one finds that they not only automatically obey the adjacency condition but in fact they obey the stronger condition of always being \emph{connected} pairs. In other words they appear together in some cluster connected by an arrow. Thus it is really the Poisson structure (i.e. the adjacency matrix $b_{ij}$) which encodes the integrability conditions. 

The cluster adjacency criterion is therefore really a constraint on the \emph{symmetric} and trivially integrable part. It implies that, even though any symmetric pair $[a \,{\scriptstyle \otimes}\, b] + [b \,{\scriptstyle \otimes}\, a]$ is integrable, to be admissible $a$ and $b$ must still appear together in some cluster. Moreover admissible pairs which are never connected by an arrow in any cluster must appear symetrically.

{\bf Triplets.}
We can go further and construct all weight three integrable words (again no restriction on initial entries) of which we find 4906. Examining the triplets therein we find two cases. Either there is a cluster in which all three letters of the triplet appear, or the triplet is of the form $[a \, {\scriptstyle \otimes}\, b \, {\scriptstyle \otimes}\, a']$ where $a'$ is the result of applying a mutation to $a$ and where $b$ is in the intersection of the connected sets of $a$ and $a'$. This implies that some pairs of letters such as $a_{11}$ and $a_{61}$ which neither share a cluster nor mutate into each other never appear together in a triplet - there must be at least two letters appearing between them in an admissible symbol.

{\bf Cluster heptagon functions.}
Now let us impose the initial entry conditions appropriate for scattering amplitudes. In the heptagon case this corresponds to allowing only $a_{1i}$ letters in the initial entries. In Table \ref{Tab:dimensions} we record the dimensions of the spaces of cluster adjacent symbols. Up to weight three all Steinmann symbols are cluster adjacent. At weight four there are 14 Steinmann symbols which fail to be cluster adjacent. They are of the form $[u\, {\scriptstyle \otimes}\, (1-u) \,{\scriptstyle \otimes}\, u \, {\scriptstyle \otimes}\, u]$ for $u = \tfrac{a_{11}a_{12}}{a_{15}}$ or $u= a_{11} a_{13}$ and cyclic copies. The failure of adjacency comes in the last two slots as in either case the letter $u$ may not appear next to itself. Interestingly, if we apply just the Steinmann conditions on the $a_{1i}$,  but everywhere in the symbol as in \cite{DP,Yorgosslides}, then up to weight six the dimensions come out to be the same as with the cluster condition i.e. the `extended' Steinmann condition and initial entry condition imply cluster adjacency. We do not yet know if this pattern continues.
\begin{center}
    \begin{table}[H]
      \begin{tabular}{r||c|c|c|c|c|c|c|c}
        Function space&\begin{minipage}{0.6cm}\centering 1\end{minipage}
        &\begin{minipage}{0.6cm}\centering 2\end{minipage}
        &\begin{minipage}{0.6cm}\centering 3\end{minipage}
        &\begin{minipage}{0.6cm}\centering 4\end{minipage}
        &\begin{minipage}{0.6cm}\centering 5\end{minipage}
        &\begin{minipage}{0.6cm}\centering 6\end{minipage}
        &\begin{minipage}{0.6cm}\centering 7\end{minipage}
        &\begin{minipage}{0.6cm}\centering 8\end{minipage}\\
        \hline\hline
        7gon& 7&42&237&1288&6763&?&?&?\\
        Steinmann 7gon& 7&28&97&322&1030&3192&9570&?\\
        Cluster 7gon& 7&28&97&308&911&2555&6826&?
      \end{tabular}
      \caption{Dimensions of various spaces constructed from the $\cal{A}$-coordinates of the $\text{Gr}(4,7)$ cluster algebra.}
      \label{Tab:dimensions}
    \end{table}
  \end{center}

 {\bf Hexagon case.}
Reducing to the hexagon case the cluster adjacency property we have described almost disappears. The only constraints are that the 
Pl\"uckers  $(i\,i+1\,i+3\,i+4)$ and $(i+1\, i+2\, i+4\, i+5)$ may not appear next to each other in the symbol which is the same as the Steinmann condition of \cite{Caron-Huot:2016owq} extended as in \cite{DP,Yorgosslides}. This is why we have focussed heavily on the heptagon case for illustration.

{\bf NMHV final entries.}  In \cite{Dixon:2016nkn} a set of final entry conditions for NMHV heptagon amplitudes due to \mbox{Caron-Huot} were given. Here we note that our adjacency condition (interpreted differentially) also applies to these combinations in the sense that the final letters of the NMHV symbols in fact lie in the same clusters as the poles of the accompanying rational functions or \emph{R-invariants}. A similar structure is exhibited in the NMHV hexagon amplitudes, as can been seen by inspecting the coproduct relations found in \cite{Dixon:2014iba}, and conjecturally will be applicable to multi-point NMHV amplitudes.  
  
{\bf  $A_2$ and $A_3$ functions.} In \cite{Golden:2014xqa} special weight-four functions were defined which have simple $B_3 {\scriptstyle \otimes} \mathbb{C}^*$ and $B_2 {\scriptstyle \wedge}  B_2$ components when written in terms of cluster $\mathcal{X}$-coordinates. The two-loop MHV amplitudes are expressible in terms of these functions modulo classical ${\rm Li}_4$ terms and products. On the other hand the patterns expressed here are most naturally described in terms of $\mathcal{A}$-coordinates. It would be fascinating to explore the connections of \cite{Golden:2014xqa} to the cluster adjacency we find here.

\section*{Acknowledgments}
\label{sec:acknowledgments}
We would like to thank Lance Dixon and Georgios Papathansiou for discussions of hexagons and Steinmann relations which inspired us to look at the singularity structure of heptagon integrals. We also thank them for encouraging us to revisit the spaces obeying extended Steinmann conditions and for independent confirmation of their dimensions. The authors are supported by ERC grant 648630 IQFT.


\bibliography{biblio}

\begin{thebibliography}{28}%
\makeatletter
\providecommand \@ifxundefined [1]{%
 \@ifx{#1\undefined}
}%
\providecommand \@ifnum [1]{%
 \ifnum #1\expandafter \@firstoftwo
 \else \expandafter \@secondoftwo
 \fi
}%
\providecommand \@ifx [1]{%
 \ifx #1\expandafter \@firstoftwo
 \else \expandafter \@secondoftwo
 \fi
}%
\providecommand \natexlab [1]{#1}%
\providecommand \enquote  [1]{``#1''}%
\providecommand \bibnamefont  [1]{#1}%
\providecommand \bibfnamefont [1]{#1}%
\providecommand \citenamefont [1]{#1}%
\providecommand \href@noop [0]{\@secondoftwo}%
\providecommand \href [0]{\begingroup \@sanitize@url \@href}%
\providecommand \@href[1]{\@@startlink{#1}\@@href}%
\providecommand \@@href[1]{\endgroup#1\@@endlink}%
\providecommand \@sanitize@url [0]{\catcode `\\12\catcode `\$12\catcode
  `\&12\catcode `\#12\catcode `\^12\catcode `\_12\catcode `\%12\relax}%
\providecommand \@@startlink[1]{}%
\providecommand \@@endlink[0]{}%
\providecommand \url  [0]{\begingroup\@sanitize@url \@url }%
\providecommand \@url [1]{\endgroup\@href {#1}{\urlprefix }}%
\providecommand \urlprefix  [0]{URL }%
\providecommand \Eprint [0]{\href }%
\providecommand \doibase [0]{http://dx.doi.org/}%
\providecommand \selectlanguage [0]{\@gobble}%
\providecommand \bibinfo  [0]{\@secondoftwo}%
\providecommand \bibfield  [0]{\@secondoftwo}%
\providecommand \translation [1]{[#1]}%
\providecommand \BibitemOpen [0]{}%
\providecommand \bibitemStop [0]{}%
\providecommand \bibitemNoStop [0]{.\EOS\space}%
\providecommand \EOS [0]{\spacefactor3000\relax}%
\providecommand \BibitemShut  [1]{\csname bibitem#1\endcsname}%
\let\auto@bib@innerbib\@empty
\bibitem [{\citenamefont {Golden}\ \emph
  {et~al.}(2014{\natexlab{a}})\citenamefont {Golden}, \citenamefont
  {Goncharov}, \citenamefont {Spradlin}, \citenamefont {Vergu},\ and\
  \citenamefont {Volovich}}]{Golden:2013xva}%
  \BibitemOpen
  \bibfield  {author} {\bibinfo {author} {\bibfnamefont {J.}~\bibnamefont
  {Golden}}, \bibinfo {author} {\bibfnamefont {A.~B.}\ \bibnamefont
  {Goncharov}}, \bibinfo {author} {\bibfnamefont {M.}~\bibnamefont {Spradlin}},
  \bibinfo {author} {\bibfnamefont {C.}~\bibnamefont {Vergu}}, \ and\ \bibinfo
  {author} {\bibfnamefont {A.}~\bibnamefont {Volovich}},\ }\href {\doibase
  10.1007/JHEP01(2014)091} {\bibfield  {journal} {\bibinfo  {journal} {JHEP}\
  }\textbf {\bibinfo {volume} {1401}},\ \bibinfo {pages} {091} (\bibinfo {year}
  {2014}{\natexlab{a}})},\ \Eprint {http://arxiv.org/abs/1305.1617}
  {arXiv:1305.1617 [hep-th]} \BibitemShut {NoStop}%
\bibitem [{\citenamefont {Hodges}(2013)}]{Hodges:2009hk}%
  \BibitemOpen
  \bibfield  {author} {\bibinfo {author} {\bibfnamefont {A.}~\bibnamefont
  {Hodges}},\ }\href {\doibase 10.1007/JHEP05(2013)135} {\bibfield  {journal}
  {\bibinfo  {journal} {JHEP}\ }\textbf {\bibinfo {volume} {1305}},\ \bibinfo
  {pages} {135} (\bibinfo {year} {2013})},\ \Eprint
  {http://arxiv.org/abs/0905.1473} {arXiv:0905.1473 [hep-th]} \BibitemShut
  {NoStop}%
\bibitem [{\citenamefont {Alday}\ and\ \citenamefont
  {Maldacena}(2007)}]{Alday:2007hr}%
  \BibitemOpen
  \bibfield  {author} {\bibinfo {author} {\bibfnamefont {L.~F.}\ \bibnamefont
  {Alday}}\ and\ \bibinfo {author} {\bibfnamefont {J.~M.}\ \bibnamefont
  {Maldacena}},\ }\href {\doibase 10.1088/1126-6708/2007/06/064} {\bibfield
  {journal} {\bibinfo  {journal} {JHEP}\ }\textbf {\bibinfo {volume} {0706}},\
  \bibinfo {pages} {064} (\bibinfo {year} {2007})},\ \Eprint
  {http://arxiv.org/abs/0705.0303} {arXiv:0705.0303 [hep-th]} \BibitemShut
  {NoStop}%
\bibitem [{\citenamefont {Drummond}\ \emph {et~al.}(2008)\citenamefont
  {Drummond}, \citenamefont {Korchemsky},\ and\ \citenamefont
  {Sokatchev}}]{Drummond:2007aua}%
  \BibitemOpen
  \bibfield  {author} {\bibinfo {author} {\bibfnamefont {J.}~\bibnamefont
  {Drummond}}, \bibinfo {author} {\bibfnamefont {G.}~\bibnamefont
  {Korchemsky}}, \ and\ \bibinfo {author} {\bibfnamefont {E.}~\bibnamefont
  {Sokatchev}},\ }\href {\doibase 10.1016/j.nuclphysb.2007.11.041} {\bibfield
  {journal} {\bibinfo  {journal} {Nucl.Phys.}\ }\textbf {\bibinfo {volume}
  {B795}},\ \bibinfo {pages} {385} (\bibinfo {year} {2008})},\ \Eprint
  {http://arxiv.org/abs/0707.0243} {arXiv:0707.0243 [hep-th]} \BibitemShut
  {NoStop}%
\bibitem [{\citenamefont {Brandhuber}\ \emph {et~al.}(2008)\citenamefont
  {Brandhuber}, \citenamefont {Heslop},\ and\ \citenamefont
  {Travaglini}}]{Brandhuber:2007yx}%
  \BibitemOpen
  \bibfield  {author} {\bibinfo {author} {\bibfnamefont {A.}~\bibnamefont
  {Brandhuber}}, \bibinfo {author} {\bibfnamefont {P.}~\bibnamefont {Heslop}},
  \ and\ \bibinfo {author} {\bibfnamefont {G.}~\bibnamefont {Travaglini}},\
  }\href {\doibase 10.1016/j.nuclphysb.2007.11.002} {\bibfield  {journal}
  {\bibinfo  {journal} {Nucl.Phys.}\ }\textbf {\bibinfo {volume} {B794}},\
  \bibinfo {pages} {231} (\bibinfo {year} {2008})},\ \Eprint
  {http://arxiv.org/abs/0707.1153} {arXiv:0707.1153 [hep-th]} \BibitemShut
  {NoStop}%
\bibitem [{\citenamefont {Steinmann}(1960{\natexlab{a}})}]{Steinmann}%
  \BibitemOpen
  \bibfield  {author} {\bibinfo {author} {\bibfnamefont {O.}~\bibnamefont
  {Steinmann}},\ }\href@noop {} {\bibfield  {journal} {\bibinfo  {journal}
  {Helv. Physica Acta}\ }\textbf {\bibinfo {volume} {33}},\ \bibinfo {pages}
  {257} (\bibinfo {year} {1960}{\natexlab{a}})}\BibitemShut {NoStop}%
\bibitem [{\citenamefont {Steinmann}(1960{\natexlab{b}})}]{Steinmann2}%
  \BibitemOpen
  \bibfield  {author} {\bibinfo {author} {\bibfnamefont {O.}~\bibnamefont
  {Steinmann}},\ }\href@noop {} {\bibfield  {journal} {\bibinfo  {journal}
  {Helv. Physica Acta}\ }\textbf {\bibinfo {volume} {33}},\ \bibinfo {pages}
  {347} (\bibinfo {year} {1960}{\natexlab{b}})}\BibitemShut {NoStop}%
\bibitem [{\citenamefont {Bartels}\ \emph {et~al.}(2010)\citenamefont
  {Bartels}, \citenamefont {Lipatov},\ and\ \citenamefont
  {Sabio~Vera}}]{Bartels:2008sc}%
  \BibitemOpen
  \bibfield  {author} {\bibinfo {author} {\bibfnamefont {J.}~\bibnamefont
  {Bartels}}, \bibinfo {author} {\bibfnamefont {L.}~\bibnamefont {Lipatov}}, \
  and\ \bibinfo {author} {\bibfnamefont {A.}~\bibnamefont {Sabio~Vera}},\
  }\href {\doibase 10.1140/epjc/s10052-009-1218-5} {\bibfield  {journal}
  {\bibinfo  {journal} {Eur.Phys.J.}\ }\textbf {\bibinfo {volume} {C65}},\
  \bibinfo {pages} {587} (\bibinfo {year} {2010})},\ \Eprint
  {http://arxiv.org/abs/0807.0894} {arXiv:0807.0894 [hep-th]} \BibitemShut
  {NoStop}%
\bibitem [{\citenamefont {Caron-Huot}\ \emph {et~al.}(2016)\citenamefont
  {Caron-Huot}, \citenamefont {Dixon}, \citenamefont {McLeod},\ and\
  \citenamefont {von Hippel}}]{Caron-Huot:2016owq}%
  \BibitemOpen
  \bibfield  {author} {\bibinfo {author} {\bibfnamefont {S.}~\bibnamefont
  {Caron-Huot}}, \bibinfo {author} {\bibfnamefont {L.~J.}\ \bibnamefont
  {Dixon}}, \bibinfo {author} {\bibfnamefont {A.}~\bibnamefont {McLeod}}, \
  and\ \bibinfo {author} {\bibfnamefont {M.}~\bibnamefont {von Hippel}},\
  }\href {\doibase 10.1103/PhysRevLett.117.241601} {\bibfield  {journal}
  {\bibinfo  {journal} {Phys. Rev. Lett.}\ }\textbf {\bibinfo {volume} {117}},\
  \bibinfo {pages} {241601} (\bibinfo {year} {2016})},\ \Eprint
  {http://arxiv.org/abs/1609.00669} {arXiv:1609.00669 [hep-th]} \BibitemShut
  {NoStop}%
\bibitem [{\citenamefont {Dixon}\ \emph {et~al.}(2017)\citenamefont {Dixon},
  \citenamefont {Drummond}, \citenamefont {Harrington}, \citenamefont {McLeod},
  \citenamefont {Papathanasiou},\ and\ \citenamefont
  {Spradlin}}]{Dixon:2016nkn}%
  \BibitemOpen
  \bibfield  {author} {\bibinfo {author} {\bibfnamefont {L.~J.}\ \bibnamefont
  {Dixon}}, \bibinfo {author} {\bibfnamefont {J.}~\bibnamefont {Drummond}},
  \bibinfo {author} {\bibfnamefont {T.}~\bibnamefont {Harrington}}, \bibinfo
  {author} {\bibfnamefont {A.~J.}\ \bibnamefont {McLeod}}, \bibinfo {author}
  {\bibfnamefont {G.}~\bibnamefont {Papathanasiou}}, \ and\ \bibinfo {author}
  {\bibfnamefont {M.}~\bibnamefont {Spradlin}},\ }\href {\doibase
  10.1007/JHEP02(2017)137} {\bibfield  {journal} {\bibinfo  {journal} {JHEP}\
  }\textbf {\bibinfo {volume} {02}},\ \bibinfo {pages} {137} (\bibinfo {year}
  {2017})},\ \Eprint {http://arxiv.org/abs/1612.08976} {arXiv:1612.08976
  [hep-th]} \BibitemShut {NoStop}%
\bibitem [{\citenamefont {Alday}\ \emph {et~al.}(2011)\citenamefont {Alday},
  \citenamefont {Gaiotto},\ and\ \citenamefont {Maldacena}}]{Alday:2009dv}%
  \BibitemOpen
  \bibfield  {author} {\bibinfo {author} {\bibfnamefont {L.~F.}\ \bibnamefont
  {Alday}}, \bibinfo {author} {\bibfnamefont {D.}~\bibnamefont {Gaiotto}}, \
  and\ \bibinfo {author} {\bibfnamefont {J.}~\bibnamefont {Maldacena}},\ }\href
  {\doibase 10.1007/JHEP09(2011)032} {\bibfield  {journal} {\bibinfo  {journal}
  {JHEP}\ }\textbf {\bibinfo {volume} {09}},\ \bibinfo {pages} {032} (\bibinfo
  {year} {2011})},\ \Eprint {http://arxiv.org/abs/0911.4708} {arXiv:0911.4708
  [hep-th]} \BibitemShut {NoStop}%
\bibitem [{\citenamefont {Goncharov}\ \emph {et~al.}(2010)\citenamefont
  {Goncharov}, \citenamefont {Spradlin}, \citenamefont {Vergu},\ and\
  \citenamefont {Volovich}}]{Goncharov:2010jf}%
  \BibitemOpen
  \bibfield  {author} {\bibinfo {author} {\bibfnamefont {A.~B.}\ \bibnamefont
  {Goncharov}}, \bibinfo {author} {\bibfnamefont {M.}~\bibnamefont {Spradlin}},
  \bibinfo {author} {\bibfnamefont {C.}~\bibnamefont {Vergu}}, \ and\ \bibinfo
  {author} {\bibfnamefont {A.}~\bibnamefont {Volovich}},\ }\href {\doibase
  10.1103/PhysRevLett.105.151605} {\bibfield  {journal} {\bibinfo  {journal}
  {Phys.Rev.Lett.}\ }\textbf {\bibinfo {volume} {105}},\ \bibinfo {pages}
  {151605} (\bibinfo {year} {2010})},\ \Eprint {http://arxiv.org/abs/1006.5703}
  {arXiv:1006.5703 [hep-th]} \BibitemShut {NoStop}%
\bibitem [{\citenamefont {Bern}\ \emph {et~al.}(2008)\citenamefont {Bern},
  \citenamefont {Dixon}, \citenamefont {Kosower}, \citenamefont {Roiban},
  \citenamefont {Spradlin} \emph {et~al.}}]{Bern:2008ap}%
  \BibitemOpen
  \bibfield  {author} {\bibinfo {author} {\bibfnamefont {Z.}~\bibnamefont
  {Bern}}, \bibinfo {author} {\bibfnamefont {L.}~\bibnamefont {Dixon}},
  \bibinfo {author} {\bibfnamefont {D.}~\bibnamefont {Kosower}}, \bibinfo
  {author} {\bibfnamefont {R.}~\bibnamefont {Roiban}}, \bibinfo {author}
  {\bibfnamefont {M.}~\bibnamefont {Spradlin}},  \emph {et~al.},\ }\href
  {\doibase 10.1103/PhysRevD.78.045007} {\bibfield  {journal} {\bibinfo
  {journal} {Phys.Rev.}\ }\textbf {\bibinfo {volume} {D78}},\ \bibinfo {pages}
  {045007} (\bibinfo {year} {2008})},\ \Eprint {http://arxiv.org/abs/0803.1465}
  {arXiv:0803.1465 [hep-th]} \BibitemShut {NoStop}%
\bibitem [{\citenamefont {Drummond}\ \emph {et~al.}(2009)\citenamefont
  {Drummond}, \citenamefont {Henn}, \citenamefont {Korchemsky},\ and\
  \citenamefont {Sokatchev}}]{Drummond:2008aq}%
  \BibitemOpen
  \bibfield  {author} {\bibinfo {author} {\bibfnamefont {J.}~\bibnamefont
  {Drummond}}, \bibinfo {author} {\bibfnamefont {J.}~\bibnamefont {Henn}},
  \bibinfo {author} {\bibfnamefont {G.}~\bibnamefont {Korchemsky}}, \ and\
  \bibinfo {author} {\bibfnamefont {E.}~\bibnamefont {Sokatchev}},\ }\href
  {\doibase 10.1016/j.nuclphysb.2009.02.015} {\bibfield  {journal} {\bibinfo
  {journal} {Nucl.Phys.}\ }\textbf {\bibinfo {volume} {B815}},\ \bibinfo
  {pages} {142} (\bibinfo {year} {2009})},\ \Eprint
  {http://arxiv.org/abs/0803.1466} {arXiv:0803.1466 [hep-th]} \BibitemShut
  {NoStop}%
\bibitem [{\citenamefont {Del~Duca}\ \emph {et~al.}(2010)\citenamefont
  {Del~Duca}, \citenamefont {Duhr},\ and\ \citenamefont
  {Smirnov}}]{DelDuca:2009au}%
  \BibitemOpen
  \bibfield  {author} {\bibinfo {author} {\bibfnamefont {V.}~\bibnamefont
  {Del~Duca}}, \bibinfo {author} {\bibfnamefont {C.}~\bibnamefont {Duhr}}, \
  and\ \bibinfo {author} {\bibfnamefont {V.~A.}\ \bibnamefont {Smirnov}},\
  }\href {\doibase 10.1007/JHEP03(2010)099} {\bibfield  {journal} {\bibinfo
  {journal} {JHEP}\ }\textbf {\bibinfo {volume} {1003}},\ \bibinfo {pages}
  {099} (\bibinfo {year} {2010})},\ \Eprint {http://arxiv.org/abs/0911.5332}
  {arXiv:0911.5332 [hep-ph]} \BibitemShut {NoStop}%
\bibitem [{\citenamefont {Fomin}\ and\ \citenamefont
  {Zelevinsky}(2002)}]{1021.16017}%
  \BibitemOpen
  \bibfield  {author} {\bibinfo {author} {\bibfnamefont {S.}~\bibnamefont
  {Fomin}}\ and\ \bibinfo {author} {\bibfnamefont {A.}~\bibnamefont
  {Zelevinsky}},\ }\href {\doibase 10.1090/S0894-0347-01-00385-X} {\bibfield
  {journal} {\bibinfo  {journal} {J. Am. Math. Soc.}\ }\textbf {\bibinfo
  {volume} {15}},\ \bibinfo {pages} {497} (\bibinfo {year} {2002})},\ \Eprint
  {http://arxiv.org/abs/math/0104151} {arXiv:math/0104151 [math.RT]}
  \BibitemShut {NoStop}%
\bibitem [{\citenamefont {Fomin}\ and\ \citenamefont
  {Zelevinsky}(2003)}]{1054.17024}%
  \BibitemOpen
  \bibfield  {author} {\bibinfo {author} {\bibfnamefont {S.}~\bibnamefont
  {Fomin}}\ and\ \bibinfo {author} {\bibfnamefont {A.}~\bibnamefont
  {Zelevinsky}},\ }\href {\doibase 10.1007/s00222-003-0302-y} {\bibfield
  {journal} {\bibinfo  {journal} {Invent. Math.}\ }\textbf {\bibinfo {volume}
  {154}},\ \bibinfo {pages} {63} (\bibinfo {year} {2003})},\ \Eprint
  {http://arxiv.org/abs/math/0208229} {arXiv:math/0208229 [math.RA]}
  \BibitemShut {NoStop}%
\bibitem [{\citenamefont {Scott}(2006)}]{Scott}%
  \BibitemOpen
  \bibfield  {author} {\bibinfo {author} {\bibfnamefont {J.~S.}\ \bibnamefont
  {Scott}},\ }\href {\doibase 10.1112/S0024611505015571} {\bibfield  {journal}
  {\bibinfo  {journal} {Proc. Lond. Math. Soc. III Ser.}\ }\textbf {\bibinfo
  {volume} {92}},\ \bibinfo {pages} {345} (\bibinfo {year} {2006})},\ \Eprint
  {http://arxiv.org/abs/0311148} {arXiv:0311148 [math]} \BibitemShut {NoStop}%
\bibitem [{\citenamefont {Drummond}\ \emph {et~al.}(2015)\citenamefont
  {Drummond}, \citenamefont {Papathanasiou},\ and\ \citenamefont
  {Spradlin}}]{Drummond:2014ffa}%
  \BibitemOpen
  \bibfield  {author} {\bibinfo {author} {\bibfnamefont {J.~M.}\ \bibnamefont
  {Drummond}}, \bibinfo {author} {\bibfnamefont {G.}~\bibnamefont
  {Papathanasiou}}, \ and\ \bibinfo {author} {\bibfnamefont {M.}~\bibnamefont
  {Spradlin}},\ }\href {\doibase 10.1007/JHEP03(2015)072} {\bibfield  {journal}
  {\bibinfo  {journal} {JHEP}\ }\textbf {\bibinfo {volume} {03}},\ \bibinfo
  {pages} {072} (\bibinfo {year} {2015})},\ \Eprint
  {http://arxiv.org/abs/1412.3763} {arXiv:1412.3763 [hep-th]} \BibitemShut
  {NoStop}%
\bibitem [{\citenamefont {Arkani-Hamed}\ \emph
  {et~al.}(2012{\natexlab{a}})\citenamefont {Arkani-Hamed}, \citenamefont
  {Bourjaily}, \citenamefont {Cachazo}, \citenamefont {Goncharov},
  \citenamefont {Postnikov} \emph {et~al.}}]{ArkaniHamed:2012nw}%
  \BibitemOpen
  \bibfield  {author} {\bibinfo {author} {\bibfnamefont {N.}~\bibnamefont
  {Arkani-Hamed}}, \bibinfo {author} {\bibfnamefont {J.~L.}\ \bibnamefont
  {Bourjaily}}, \bibinfo {author} {\bibfnamefont {F.}~\bibnamefont {Cachazo}},
  \bibinfo {author} {\bibfnamefont {A.~B.}\ \bibnamefont {Goncharov}}, \bibinfo
  {author} {\bibfnamefont {A.}~\bibnamefont {Postnikov}},  \emph {et~al.},\
  }\href@noop {} {\  (\bibinfo {year} {2012}{\natexlab{a}})},\ \Eprint
  {http://arxiv.org/abs/1212.5605} {arXiv:1212.5605 [hep-th]} \BibitemShut
  {NoStop}%
\bibitem [{\citenamefont {Drummond}\ \emph {et~al.}(2010)\citenamefont
  {Drummond}, \citenamefont {Henn}, \citenamefont {Korchemsky},\ and\
  \citenamefont {Sokatchev}}]{Drummond:2007au}%
  \BibitemOpen
  \bibfield  {author} {\bibinfo {author} {\bibfnamefont {J.}~\bibnamefont
  {Drummond}}, \bibinfo {author} {\bibfnamefont {J.}~\bibnamefont {Henn}},
  \bibinfo {author} {\bibfnamefont {G.}~\bibnamefont {Korchemsky}}, \ and\
  \bibinfo {author} {\bibfnamefont {E.}~\bibnamefont {Sokatchev}},\ }\href
  {\doibase 10.1016/j.nuclphysb.2009.10.013} {\bibfield  {journal} {\bibinfo
  {journal} {Nucl.Phys.}\ }\textbf {\bibinfo {volume} {B826}},\ \bibinfo
  {pages} {337} (\bibinfo {year} {2010})},\ \Eprint
  {http://arxiv.org/abs/0712.1223} {arXiv:0712.1223 [hep-th]} \BibitemShut
  {NoStop}%
\bibitem [{\citenamefont {Caron-Huot}(2011)}]{CaronHuot:2011ky}%
  \BibitemOpen
  \bibfield  {author} {\bibinfo {author} {\bibfnamefont {S.}~\bibnamefont
  {Caron-Huot}},\ }\href {\doibase 10.1007/JHEP12(2011)066} {\bibfield
  {journal} {\bibinfo  {journal} {JHEP}\ }\textbf {\bibinfo {volume} {1112}},\
  \bibinfo {pages} {066} (\bibinfo {year} {2011})},\ \Eprint
  {http://arxiv.org/abs/1105.5606} {arXiv:1105.5606 [hep-th]} \BibitemShut
  {NoStop}%
\bibitem [{\citenamefont {Caron-Huot}\ \emph {et~al.}()\citenamefont
  {Caron-Huot}, \citenamefont {Dixon}, \citenamefont {McLeod}, \citenamefont
  {von Hippel},\ and\ \citenamefont {Papathanasiou}}]{DP}%
  \BibitemOpen
  \bibfield  {author} {\bibinfo {author} {\bibfnamefont {S.}~\bibnamefont
  {Caron-Huot}}, \bibinfo {author} {\bibfnamefont {L.~J.}\ \bibnamefont
  {Dixon}}, \bibinfo {author} {\bibfnamefont {A.}~\bibnamefont {McLeod}},
  \bibinfo {author} {\bibfnamefont {M.}~\bibnamefont {von Hippel}}, \ and\
  \bibinfo {author} {\bibfnamefont {G.}~\bibnamefont {Papathanasiou}},\
  }\href@noop {} {\bibinfo  {journal} {to appear}\ }\BibitemShut {NoStop}%
\bibitem [{\citenamefont {Papathanasiou}()}]{Yorgosslides}%
  \BibitemOpen
\bibfield  {journal} {  }\bibfield  {author} {\bibinfo {author} {\bibfnamefont
  {G.}~\bibnamefont {Papathanasiou}},\ }\href@noop {} {\bibinfo  {journal}
  {talk at Amplitudes 2017}\ }\BibitemShut {NoStop}%
\bibitem [{\citenamefont {Arkani-Hamed}\ \emph
  {et~al.}(2012{\natexlab{b}})\citenamefont {Arkani-Hamed}, \citenamefont
  {Bourjaily}, \citenamefont {Cachazo},\ and\ \citenamefont
  {Trnka}}]{ArkaniHamed:2010gh}%
  \BibitemOpen
\bibfield  {journal} {  }\bibfield  {author} {\bibinfo {author} {\bibfnamefont
  {N.}~\bibnamefont {Arkani-Hamed}}, \bibinfo {author} {\bibfnamefont {J.~L.}\
  \bibnamefont {Bourjaily}}, \bibinfo {author} {\bibfnamefont {F.}~\bibnamefont
  {Cachazo}}, \ and\ \bibinfo {author} {\bibfnamefont {J.}~\bibnamefont
  {Trnka}},\ }\href {\doibase 10.1007/JHEP06(2012)125} {\bibfield  {journal}
  {\bibinfo  {journal} {JHEP}\ }\textbf {\bibinfo {volume} {06}},\ \bibinfo
  {pages} {125} (\bibinfo {year} {2012}{\natexlab{b}})},\ \Eprint
  {http://arxiv.org/abs/1012.6032} {arXiv:1012.6032 [hep-th]} \BibitemShut
  {NoStop}%
\bibitem [{\citenamefont {Drummond}\ \emph {et~al.}(2011)\citenamefont
  {Drummond}, \citenamefont {Henn},\ and\ \citenamefont
  {Trnka}}]{Drummond:2010cz}%
  \BibitemOpen
  \bibfield  {author} {\bibinfo {author} {\bibfnamefont {J.~M.}\ \bibnamefont
  {Drummond}}, \bibinfo {author} {\bibfnamefont {J.~M.}\ \bibnamefont {Henn}},
  \ and\ \bibinfo {author} {\bibfnamefont {J.}~\bibnamefont {Trnka}},\ }\href
  {\doibase 10.1007/JHEP04(2011)083} {\bibfield  {journal} {\bibinfo  {journal}
  {JHEP}\ }\textbf {\bibinfo {volume} {04}},\ \bibinfo {pages} {083} (\bibinfo
  {year} {2011})},\ \Eprint {http://arxiv.org/abs/1010.3679} {arXiv:1010.3679
  [hep-th]} \BibitemShut {NoStop}%
\bibitem [{\citenamefont {Dixon}\ and\ \citenamefont {von
  Hippel}(2014)}]{Dixon:2014iba}%
  \BibitemOpen
  \bibfield  {author} {\bibinfo {author} {\bibfnamefont {L.~J.}\ \bibnamefont
  {Dixon}}\ and\ \bibinfo {author} {\bibfnamefont {M.}~\bibnamefont {von
  Hippel}},\ }\href {\doibase 10.1007/JHEP10(2014)065} {\bibfield  {journal}
  {\bibinfo  {journal} {JHEP}\ }\textbf {\bibinfo {volume} {1410}},\ \bibinfo
  {pages} {65} (\bibinfo {year} {2014})},\ \Eprint
  {http://arxiv.org/abs/1408.1505} {arXiv:1408.1505 [hep-th]} \BibitemShut
  {NoStop}%
\bibitem [{\citenamefont {Golden}\ \emph
  {et~al.}(2014{\natexlab{b}})\citenamefont {Golden}, \citenamefont {Paulos},
  \citenamefont {Spradlin},\ and\ \citenamefont {Volovich}}]{Golden:2014xqa}%
  \BibitemOpen
  \bibfield  {author} {\bibinfo {author} {\bibfnamefont {J.}~\bibnamefont
  {Golden}}, \bibinfo {author} {\bibfnamefont {M.~F.}\ \bibnamefont {Paulos}},
  \bibinfo {author} {\bibfnamefont {M.}~\bibnamefont {Spradlin}}, \ and\
  \bibinfo {author} {\bibfnamefont {A.}~\bibnamefont {Volovich}},\ }\href
  {\doibase 10.1088/1751-8113/47/47/474005} {\bibfield  {journal} {\bibinfo
  {journal} {J. Phys.}\ }\textbf {\bibinfo {volume} {A47}},\ \bibinfo {pages}
  {474005} (\bibinfo {year} {2014}{\natexlab{b}})},\ \Eprint
  {http://arxiv.org/abs/1401.6446} {arXiv:1401.6446 [hep-th]} \BibitemShut
  {NoStop}%
\end{thebibliography}%

\end{document}